\documentstyle[aps,multicol]{revtex}
\input epsf
\begin{document}
\draft
\title{Power-Law Sensitivity to Initial Conditions within a Logistic-like
Family of Maps: Fractality and Nonextensivity}   
\author{U.M.S. Costa, M.L.Lyra,}\address{Departamento de F\'{\i}sica, 
Universidade Federal de Alagoas, Maceio-AL, Brazil}
\author{A.R. Plastino 
and C. Tsallis} \address{Centro Brasileiro de Pesquisas F\'{\i}sicas\\
Rua Xavier Sigaud 150, 22290-180 -- Rio de Janeiro -- RJ, Brazil \\
e-mail: tsallis@cat.cbpf.br}

\maketitle

\begin{abstract}
 Power-law sensitivity to initial conditions, characterizing the
behaviour of dynamical systems at their critical points (where the
standard Liapunov exponent vanishes), is studied in connection with the
family of nonlinear 1D logistic-like maps
$x_{t+1} \, = \, 1 \, -
\, a \, \left | x_t \right |^z, \, (z > 1; 0<a\le 2; t=0,1,2, \ldots )$. The main
ingredient of our approach is the generalized deviation law $\lim_{\Delta
x(0) \rightarrow 0} \frac{\Delta x(t)}{\Delta
x(0)}=[1+(1-q)\lambda_q\;t]^{\frac{1}{1-q}}$ (equal to $e^{\lambda_1 t}$
for $q=1$, and proportional, for large $t$, to $t^{\frac{1}{1-q}}$ for 
$q\ne1$; $q \in  \cal{R}$ is the entropic index appearing in the recently 
introduced nonextensive generalized statistics).  The relation between the 
parameter $q$ and the fractal dimension $d_f$ of the
onset-to-chaos attractor  is revealed: $q$ appears to monotonically decrease 
from 1 (Boltzmann-Gibbs, extensive, limit) to $-\infty$ when $d_f$ varies 
from  1 (nonfractal, ergodic-like, limit) to zero.  

\end{abstract}
\begin{multicols}{2}
\narrowtext

\section{Introduction}

        The standard thermostatistical formalism of Boltzmann-Gibbs (BG)
constitutes one of the most successful paradigms of theoretical physics.
It provides the link between microscopic dynamics and the macroscopic
properties of matter. Inspired in Shannon's Information Theory
\cite{chinchin}, Jaynes \cite{jaynes}
reformulation of the BG theory greatly increased its power and scope.
Jaynes provided a general prescription for the construction of a
probability distribution $f({\bf x})$ (${\bf x} \in R^d $ stands for a
point in the relevant phase space), when the only available information
about the system are the mean values of $M$ quantities

\begin{equation} \label{mean}
\langle A_r ({\bf x}) \rangle \, \equiv \, \int \, A_r ({\bf x}) \, f({\bf x})\, 
d{\bf x}, \,\,\,\,\,\,\, (r=1, \ldots ,M).
\end{equation}

\noindent
According to Jaynes, the least biased distribution compatible with the
data (\ref{mean}) is the one that maximizes Shannon's Information,

\begin{equation} \label{slog}
S_1 \, \equiv \, -\int \, f({\bf x})\, \ln f({\bf x})\, d{\bf x}, 
\end{equation}
(the use of the subindex $1$ will become transparent later on) under 
the constraints imposed by the mean values (\ref{mean}) and
appropriate normalization 

\begin{equation} \label{norm}
\int \, f({\bf x})\, d{\bf x} \, = \,1.
\end{equation}

The well known answer to the above variational problem is provided by the
maximum entropy (ME) distribution

\begin{equation} \label{maxent}
f_{ME}({\bf x})\, = \, \frac{1}{Z_1} \, 
\exp \left(-\sum_{r=1}^M \, \lambda_r A_r ({\bf x}) \right) ,
\end{equation}

\noindent
where $\{\lambda_r\}$ are the $M$ Lagrange multipliers associated with the
known mean values, and the partition function $Z_1$ is given by 

\begin{equation} \label{z}
Z_1 \, \equiv \, \int \,  
\exp \left(-\sum_{r=1}^M \, \lambda_r A_r({\bf x}) \right) \, d{\bf x}.
\end{equation}

Jaynes' prescription can be regarded as a mathematical formulation of the
celebrated ``Occam's Razor" principle. In order to obtain a statistical
description of a system, given by the distribution $f({\bf x})$, we must
employ all and only the disponible data (\ref{mean}), without assuming any further
information we do not actually have. 

Jaynes informational approach allows to consider more general statistical
ensembles than the Gibbs microcanonical, canonical, and macrocanonical.
Also it provides a natural way to treat nonequilibrium situations.

Despite its great success, the Boltzmann-Gibbs-Jaynes formalism is unable
to deal with a variety of interesting physical problems such as the
thermodynamics of self-gravitating systems, some anomalous diffusion 
phenomena,
L\'evy flights and distributions, turbulence, among others (see \cite{tsallis2} for a
more detailed list). In order to deal with these difficulties, Jaynes
approach is compatible with exploring the possibility of building up a 
thermostatistics
based upon an entropy functional different from the usual logarithmic
entropy. Recently one of us introduced \cite{tsallis} the following
generalized, nonextensive entropy form    

\begin{equation} \label{sq}
S_q \, \equiv \,  \frac {1 \, - \, \int [f({\bf x})]^q d{\bf x}}{q \, - \, 1},
\end{equation}

\noindent
where $q$ is a real parameter characterizing the entropy functional $S_q$.
This entropy recovers $S_1$ as the $q=1$ particular instance and was 
introduced in order to describe systems where
nonextensivity plays an important role; if $A$ and $B$ are two independent 
systems (in the sense that the probabilities associated with $A+B$ factorize 
into those of $A$ and $B$) we straightforwardly verify that $S_q(A+B)=
S_q(A)+S_q(B)+(1-q)\,S_q(A)\,S_q(B) $. Indeed, nonextensive
behaviour is the common feature among the above listed problems where the
usual statistics fails.  The generalized nonextensive thermostatistics
has already been applied to astrophysical self-gravitating systems
\cite{plastino}, the solar neutrino problem \cite{kaniadakis}, distribution of 
peculiar velocities of galaxy clusters \cite{lavagno}, cosmology \cite{hamity}, 
two-dimensional turbulence in pure-electron
plasma \cite{boghosian}, anomalous diffusions of the L\'evy \cite{levy} and 
correlated \cite{correlated} types, long-range magnetic and Lennard-Jones-like 
systems \cite{jund}, simulated annealing and other
optimization techniques \cite{optimization}, dynamical linear response
theory \cite{rajagopal}, among others.      

The nonextensivity effects displayed by the above listed systems can
arise from long-range interactions, long-range microscopic memory, or
fractal space-time constraints. Even for dynamical systems that ``live" in an
euclidean (nonfractal) space, if the subset (of this space) that the system visits 
(most of the time) during
its evolution has a fractal geometry, the generalized
thermostatistics might provide a better account of the situation than that
provided by the usual statistics. Indeed, it is well known that
nonlinear chaotic dynamical systems may have fractal attractors
\cite{chaos}. Two of 
the most important dynamical quantities usually employed in order to
characterize such chaotic systems, are the Liapunov exponents
\cite{definitions,hilborn}, and the 
Kolmogorov-Sinai (KS) entropy \cite{kolsinai}. In a recent effort
\cite{tzp} (TPZ from here on), generalizations for these quantities
inspired in the 
generalized nonextensive entropy $S_q$ (and its consequences) were 
introduced. The generalized
Liapunov exponent $\lambda_q$ and generalized KS-entropy $K_q$ provide a useful
characterization of the dynamics corresponding to critical points where
the usual Liapunov exponent vanishes. For these critical cases, the exponential
sensitivity to initial conditions is replaced by a power-law one, and the
vanishing (standard) Liapunov exponent $\lambda_1$  provides but a poor description of
the concomitant dynamics. On the contrary, the generalized exponent
$\lambda_q$ appropriately discriminates between the different possible
power-law behaviours. TPZ illustrates these ideas with the (good
old) logistic map. It is of interest to explore this formalism  as applied
to other nonlinear dynamical systems. In particular, it is of importance to study
families of dynamical systems characterized by a set of parameters. Each
member of the family will have a different onset-to-chaos critical point, with  a
corresponding attractor characterized by a Hausdorff fractal dimension 
$d_f$ and a suitable value of the entropic parameter $q$. The study of these families  
will enlighten the relation between $q$ and $d_f$. 

   The specific aim of the present paper is to study the relation between the
fractal dimension $d_f$ of the onset-to-chaos attractor and the
parameter $q$ for the logistic-like family of maps (see \cite{feigenbaum,hauser,bailin} 
and references therein)

\begin{equation} \label{za}
x_{t+1} \, = \, 1 \, - \, a \, \left | x_t \right |^z, \,\,\, 
\end{equation}

\noindent
($z > 1; 0<a\le 2; t=0,1,2, \ldots; x_t \in [-1,1] $). Note that in the particular case $z=2$ we recover the standard logistic map 
(in its centered representation). 

The paper is organized as follows. In Section II we briefly review the
$q$-generalizations of the Liapunov exponent and the KS-entropy. In
Section III the results for the logistic-like maps are presented. Our main
conclusions are drawn in Section IV.

\section{Generalized Liapunov exponent and KS-entropy.}

 Let us consider, for a one dimensional dynamical system, two nearby
orbits whose initial conditions differ by the small quantity $\Delta x(0)
$. We will assume that the time dependence of the distance between both
orbits is given by the ansatz \cite{tzp}

\begin{equation} \label{power1}
\lim_{\Delta x(0) \rightarrow 0}\frac{\Delta x(t)}{\Delta x(0)}=
[1+(1-q)\;\lambda_q\;t]^\frac{1}{1-q}\;\;\;
 (q \in {\cal R})
\end{equation}

\noindent
where $\lambda_q $ is our generalized Liapunov exponent, and $q$ is a
real parameter characterizing the behaviour of the system.
We verify that this equation is identically satisfied for $t=0$ ($\forall
q$), and that  $q  \ne 1$ yields, for large times, the {\it power-law}
 
\begin{equation} \label{power2}
\lim_{\Delta x(0) \rightarrow 0}\frac{\Delta x(t)}{\Delta x(0)} \sim
[(1-q)\lambda_q]^{\frac{1}{1-q}}\;t^{\frac{1}{1-q}} 
\,\,\,\, (t \rightarrow \infty)
\end{equation}

\noindent
On the other hand, it is plain that for $q \rightarrow 1$ we recover the
standard {\it exponential} deviation law

\begin {equation} \label{expd}
\lim_{\Delta x(0) \rightarrow 0}\frac{\Delta x(t)}{\Delta x(0)}= \exp [\lambda_1\;t] \;\;\;
\end {equation}

\noindent 
where $\lambda_1$ is just the usual Liapunov exponent.
The $q=1$ scenario corresponds to situations with $\lambda_1 \ne 0$. These
cases describe chaotic behaviour ($ \lambda_1 > 0$) and regular behaviour 
($\lambda_1 < 0$). The generalized exponent $\lambda_q$ is intended to
provide a convenient description of the marginal situations where the usual 
Liapunov
exponent vanishes ($\lambda_1=0$). In these last cases, we have the power 
law sensibility
to initial conditions given by Eq.(\ref{power2}) instead of the usual
exponential one. 
 The generalized deviation law (Eq.(\ref{power1})) is inspired in the form of the
$q$-generalized nonextensive canonical distribution, given by \cite{tsallis}

\begin{equation} \label{canonq}
p_i \, =\,\frac{[1\,-\,(1-q)\beta\,\epsilon_i ]^{1/(1-q)}}{Z_q},
\end{equation}
with the generalized partition function being given by

\begin{equation} \label{zq}
Z_q \equiv \sum_i [1\,-\,(1-q)\beta\,\epsilon_i]^{1/(1-q)}
\end{equation}
\noindent
where $\beta\equiv1/kT$ and $\{\epsilon_i\}$ is the full set of eigenvalues of 
the Hamiltonian of the system.
\noindent
Notice that, in the limit $q \rightarrow 1$, this thermal canonical 
equilibrium distribution reduces to the ordinary BG one

\begin{equation} \label{canon}
p_i \, =\frac{\exp [ - \, \beta \, \epsilon_i ]}{Z_1} 
\end{equation}
with

\begin{equation} \label{z1}
Z_1\equiv \sum_i \exp[-\,\beta\,\epsilon_i]
\end{equation}

It is worth to remark that the marginal case with vanishing (standard)
Liapunov exponent $\lambda_1$ displays a very rich and complex 
behaviour, reminiscent
of what happens at the critical point of thermal equilibrium critical
phenomena. To just say that $\lambda_1 =0$ is a very poor description of
its richness, intimately connected to fractality. Indeed, within our
generalized formalism, the parameter $q$ provides a characterization of
the kind of power-law sensitivity to initial conditions involved, and is
expected to be related to the fractal dimension $d_f$ of the corresponding
attractor.  

  In order to discuss the generalized KS-entropy, let us consider a
partition of phase space in cells with size characterized by a linear
scale $l$. We will study the evolution of an ensemble of identical copies
of our system. We assume that all the members of the ensemble start at
$t=0$ with initial conditions belonging to one and the same cell. This
means that the probability associated to that privileged cell is 1, while
the remaining cells of the partition have vanishing initial probabilities.
As time goes by, and due to the sensitivity to initial conditions, our 
ensemble will spread over an increasing number of cells. The standard
KS-entropy can be regarded as the rate of growth of the Boltzmann-Gibbs
entropy associated with the partition probability distribution.   
   
  Within the generalized nonextensive thermostatistics, the entropy
functional for a discrete probability distribution $\{ p_i \} $ is given
by 
 
\begin{equation}  \label{sqdisc}
S_q = \frac{1-\sum_{i=1}^W p_i^q}{q-1}\;\;\;(q \in \cal{R})
\end{equation}

\noindent
which, for equiprobability, becomes

\begin{equation} \label{equipro}
S_q=\frac{W^{1-q}-1}{1-q}
\end{equation}

\noindent
The use of  Eq. (\ref{sqdisc}), instead of $S_1 \,=\, -\sum_{i=1}^W \,p_i \, \ln
p_i $, yields (along Zanette's lines \cite{zanettekol}) to the following
generalization of the Kolmogorov-Sinai entropy 

\begin{equation} \label{ksentr}
K_q \equiv  \lim_{ \tau  \rightarrow  0} \lim_{l  \rightarrow  0}\lim_{N
\rightarrow  \infty} 
 \frac{1}{N \tau} \left( S_q(N) - S_q(0) \right),
 \end{equation}

\noindent
that under the assumption of equiprobability reduces to

\begin{equation} \label{kqequi}
K_q =  \lim_{ \tau  \rightarrow  0} \lim_{l  \rightarrow  0}\lim_{N
\rightarrow  \infty} 
\frac{1}{N \tau} \frac{[W(N)]^{(1-q)}-1}{1-q}. 
\end{equation}

\noindent
In both equations (\ref{ksentr},\ref{kqequi}) we have maintained 
the traditional $\tau \rightarrow 0$ which applies for a continuous 
time $t$; it is clear however that, in our present case, this limit does not 
apply since our $t$ is discrete.
We must remark that our generalization $K_q$ of the KS entropy is
different from the generalizations $K(\beta)$ based upon Renyi
informations, usually called "Renyi entropies" in the literature of
thermodynamics of chaotic systems \cite{beckscho} (sometimes the parameter
characterizing these generalizations of the KS entropies is called $q$
instead of $\beta$ \cite{grass}. This parameter $q$ should not be confused
with our $q$).

\noindent
Consistently with the behavior indicated in Eq. (\ref{power1}), we have 
(along Hilborn's lines \cite{hilborn})

\begin{equation}
W(N)=\;[1+(1-q)\;\lambda_q\;N \tau]^{\frac{1}{1-q}}
\end{equation}

\noindent
which, replaced into Eq. (\ref{kqequi}), immediately yields (for 1D dynamical
systems) 

\begin{equation}
K_q=\lambda_q
\end{equation}

This relation holds if $\lambda_q>0$ ($K_q$ vanishes if $\lambda_q \le 0$); it 
constitutes a generalization of the well known Pesin equality
$K_1=\lambda_1$ (if $\lambda_1>0$; $K_1=0$ otherwise), and unifies (within a  
single scenario for both
exponential and power-law sensitivities to initial conditions)  the
connection between sensitivity and rythm of loss of information.\\

\section{The Logistic-like Map.}

Let us now illustrate some of the above concepts by focusing the logistic-like 
maps (\ref{za}). These maps are relatively well known and have been 
addressed in various occasions (\cite{hauser,bailin} and references therein). 
The {\it topological} properties associated with them (such as the sequence of 
attractors while varying the parameter $a$) do not depend on $z$, but the 
{\it metrical} properties (such as Feigenbaum's exponents) do depend on $z$. 
We shall exhibit herein that the same occurs with $q$. Indeed, although quite 
a lot is known for these maps, their sensitivity to the initial conditions at the 
onset-to-chaos has never been addressed as far as we know. As we shall see, 
for all values of $z$, the sensitivity is of the {\it weak} type \cite{tzp}, i.e., 
power-laws instead of the usual exponential ones.

\begin{figure}
\setlength{\epsfxsize}{9.cm}
\centerline{\mbox{\epsffile{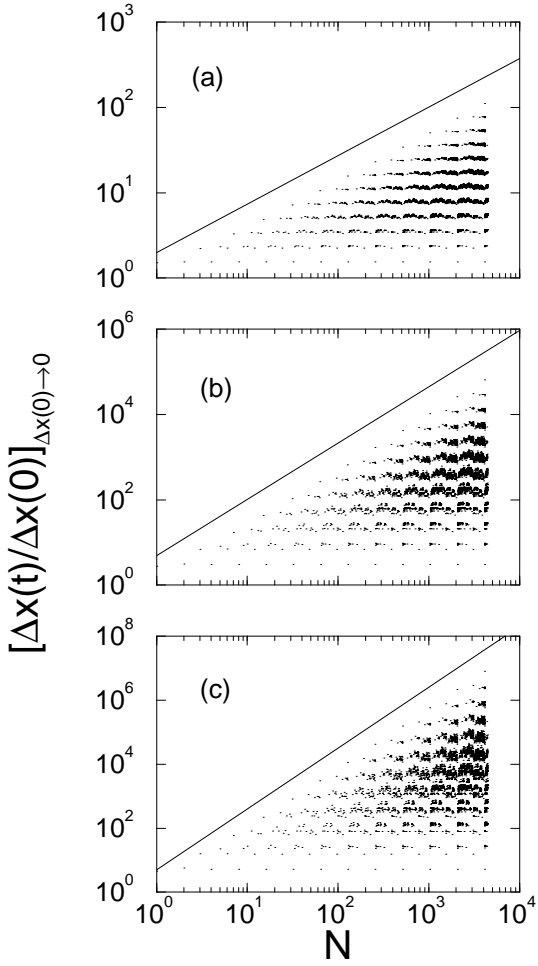}}}
\caption{Log-log plot of 
$\lim_{\Delta x(0) \rightarrow 0}\left( \frac{\Delta x(t)}{\Delta x(0)}\right)$ 
versus the number
of iterations $N$ calculated for $x_0=0$ (the slope $1/(1-q)$ is calculated 
using  the upper bound points): (a) $z=1.25$; (b) $z=2$; (c) $z=3$}
\label{fig1}
\end{figure}

We present now our main numerical results. We computed, as functions of $z$, 
the parameter $q$ and the critical fractal dimension $d_f$ (determined within 
the box counting procedure).

In Fig. 1 we exhibit, for typical values of $z$ at its chaotic threshold $a_c(z)$ 
and using 
$x_0=0$, a plot of $\,\ln \lim_{\Delta x(0) \rightarrow 0}\frac{\Delta x(t)}{\Delta x(0)}\,=
\,\sum_{n=1}^N \ln\,[a\,z\left|x_n\right|^{z-1}]$ versus  $\ln\,N$, where $N$ is 
the number of iterations. (Notice that for convenience we use, as argument of the 
logarithm, not exactly the derivative $dx_{t+1}/dx_t$ , but rather its absolute value). 
For each of these plots we see an upper bound whose slope equals $1/(1-q)$ 
(see Eq. (\ref{power2})), from which we determine $q$.

\begin{figure}
\setlength{\epsfxsize}{6.cm}
\centerline{\mbox{\epsffile{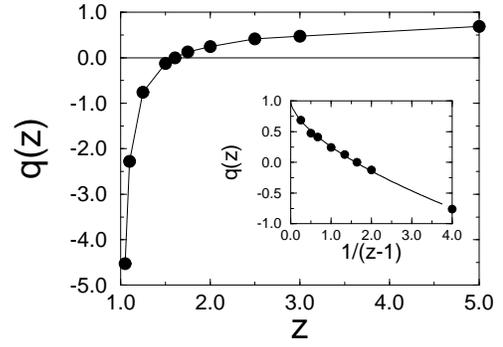}}}
\caption{$z$-dependence of the entropic index $q$. The inset conveniently 
represents the $z \rightarrow \infty$ neiborhood  (the continuous line is the 
best fitting with a curve $q=1-a_0/(z-1)^{a_1}$; we obtained  
$a_0=0.746$ and $a_1=0.613$).}
\label{fig2}
\end{figure}

In Fig. 2 we show the behaviour of the parameter $q$ as a function of the
parameter $z$ characterizing the map. The figure suggests that for
$z\rightarrow 1$, $q$ tends to $-\infty$, while in the limit $z
\rightarrow \infty$, $q$ approaches 1.
In Fig. 3 we can see, for typical values of $z$ at its chaotic threshold $a_c(z)$, 
the number of filled boxes as function of the number of boxes, corresponding to 
the box counting method employed in order to
determine the fractal dimension $d_f$.
In Fig. 4 the behaviour of the fractal dimension  $d_f$ of the chaotic critical
attractor as function of $z$ is depicted. We can observe that, as
$z\rightarrow 1$, $d_f$ seems to go to 0, while, in the limit $z
\rightarrow \infty$, the $d_f$-curve approaches unity.In Fig. 5 we show the behaviour of the parameter $q$ as a function of the
fractal dimension $d_f$. We can see that $q$ displays a monotonically increasing  behaviour with the fractal dimension $d_f$.\\
\begin{figure}
\setlength{\epsfxsize}{6.cm}
\centerline{\mbox{\epsffile{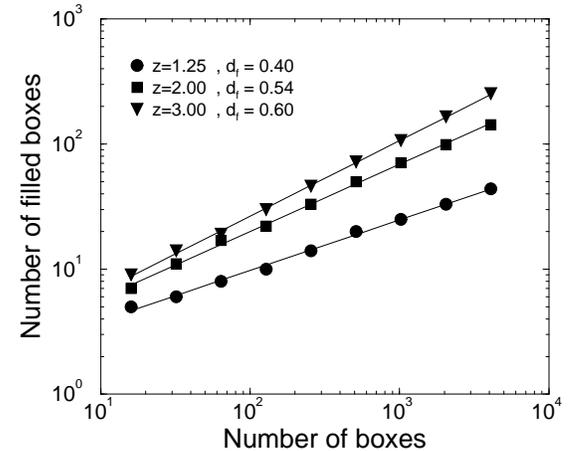}}}
\caption{Box counting procedure for determining the onset-to-chaos fractal 
dimension $d_f$ for typical values of $z$.}
\label{fig3}
\end{figure}
\begin{figure}
\setlength{\epsfxsize}{6.cm}
\centerline{\mbox{\epsffile{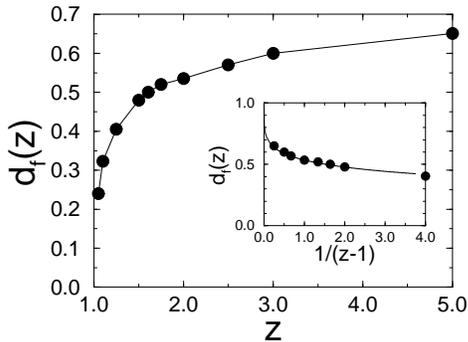}}}
\caption{ $z$-dependence of the fractal dimension $d_f$. The inset conveniently 
represents the $z \rightarrow \infty$ neiborhood  (the continuous line is the 
best fitting with a curve $d_f=\exp{-[b_0/(z-1)^{b_1}]}$; we obtained  
$b_0=0.62$ and $b_1=0.27$).}
\label{fig4}
\end{figure}
 The particular value $q=0$,
that describes {\it linear} sensitivity to initial conditions corresponds,
with notable numerical accuracy, to the fractal dimension $d_f =0.5$ 
(numerically $d_f = 0.50 \pm0.01$, occuring for $z=1.609\pm0.001$). It is remarkable that as the fractal dimension tends towards 1, the
parameter $q$ approaches 1. If this tendency becomes confirmed by 
analytic results or more powerful numerical work,  this would be very 
enlightening, because
in the limit $d_f \rightarrow 1$ the attractor loses its fractal nature (in the 
sense that $d_f$ coincides with the euclidean dimension $d=1$),
and the usual statistics (i.e., the usual exponential deviation of nearby
trajectories), characterized by $q=1$, would be recovered. On the
other extreme, as $d_f \rightarrow 0$, $q$ appears to approach $-\infty$, 
hence $1/(1-q)=0$, which can be considered as an indication of a possible 
{\it logarithmic} sensitivity to the initial conditions. Our results are 
summarized in Table I.

\begin{figure}
\setlength{\epsfxsize}{6.cm}
\centerline{\mbox{\epsffile{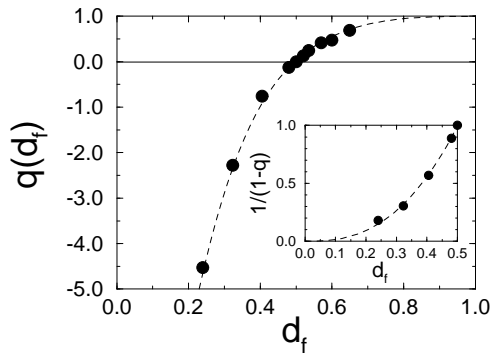}}}
\caption{$d_f$-dependence of the entropic index $q$; the inset conveniently 
represents the $d_f  \le 1/2$ region. Notice that the point $(d_f,q)=(1/2,0)$ 
seems to belong to the curve. The dashed lines are guides to the eye.}
\label{fig5}
\end{figure}

\begin{center}
\begin{tabular}{||l|l|l|l||} \hline
$~~z$ & $~~~~~~a_c$ & $~~~~~~~q$ & $~~~~~d_f$ \\ \hline
$1$ & $1^*$ & $-\infty^*$ & $0^*$ \\ \hline
$1.05$ & $1.0816488...$ & $-4.52\pm 0.03$ & $0.24\pm 0.02$ \\ \hline
$1.10$ & $1.1249885...$ & $-2.28\pm 0.02$ & $0.32 \pm 0.02$ \\ \hline
$1.25$ & $1.2095137...$ & $-0.76\pm 0.01$ & $0.40\pm 0.01$ \\ \hline
$1.5$ &  $1.2955099...$ & $-0.12\pm 0.01$ & $0.48\pm 0.01$ \\ \hline
$1.609$ & $1.3236435...$ & $~~0.00\pm 0.01$ & $0.50\pm 0.01$ \\ \hline
$1.75$ & $1.3550607...$ & $~~0.13\pm 0.01$ & $0.52\pm 0.01$ \\ \hline
$2.0$ & $ 1.4011551...$ & $~~0.24\pm 0.01$ & $0.54\pm 0.01$ \\ \hline
$2.5$ & $1.4705500...$ & $~~0.41\pm 0.01$ & $ 0.57\pm 0.01$ \\ \hline
$3.0$ & $1.5218787...$ & $~~0.47\pm 0.01$ &$ 0.60\pm 0.01$ \\ \hline
$5.0$ & $1.6456203...$ & $~~0.69\pm 0.01$ &$ 0.65\pm 0.01$ \\ \hline
$\infty $ & $2^*$ & $~~1^*$ & $1^*$ \\ \hline
\end{tabular}
\end{center}
\noindent
TABLE I - * denotes the limiting values suggested by the numerical results; 
$a_c=1$ for $z=1$ also has analytic support (see \cite{hauser} and references therein); 
$a_c=2$ for $z\rightarrow \infty$ also has renormalization group support \cite{hauser}.

\section{Conclusions.}
We have exhibited, for a family of logistic-like maps , the
behaviours of the entropic parameter $q$ and the fractal dimension $d_f$ of the
onset-to-chaos attractor. We showed that, at this critical point, {\it power} deviation
laws for nearby orbits, similar to the ones appearing \cite{tzp} in the logistic map,
are observed. The concomitant value of $q$ is related to the chaotic
attractor fractal dimension. It would no doubt be interesting to find out if, for 
generic nonlinear dynamical systems, $q$ depends only on $d_f$ (support of 
the visiting frequency function) or also upon other characteristics of the 
critical attractor, such as the
visiting frequency function itself. In order to answer
this question, it would be useful to explore the behavior of families of
maps whose possible chaotic critical points depend on more than one 
parameter. Such studies, as well as the application of these concepts to 
self-organized criticality \cite{perbak}, would be very welcome.\\
Finally, let us stress that the present study provides a direct and important 
insight onto a problem which has been quite ellusive up to now, namely the 
microscopic interpretation of the entropic index $q$ characterizing 
nonextensive statistics. 
The present results clearly exhibit that what determines $q$ is not the 
entire phase space within which the system is allowed to evolve (the 
euclidean interval $-1 \le x_t \le 1$ in the present examples), but the 
(possibly fractal) subset of it onto which the system is driven by its 
own dynamics. Consistently, whenever  the relevant fractal dimension 
approaches its associated euclidean value ($d=1$ in the present case), 
extensivity (i.e., $q=1$) and standard BG thermostatistics naturally 
become, as is well known, the appropriate standpoints.\\

\end{multicols}

\begin{references}
\bibitem{chinchin} Khinchin, A.I., {\it Mathematical Fundations of 
Information Theory}, (Dover Publ., New York, 1957).

\bibitem{jaynes} E.T. Jaynes in {\it Statistical Physics}, ed. W.K. Ford 
(Benjamin, New York, 1967). 

\bibitem{tsallis2}C. Tsallis, Fractals {\bf 3}, 541(1995).

\bibitem{tsallis} C. Tsallis, J. Stat. Phys. {\bf 52}, 479 (1988); E.M.F.
Curado and C. Tsallis, J. Phys. A {\bf 24}, L69 (1991)[Corrigenda: {\bf
24}, 3187 (1991) and {\bf 25}, 1019 (1992)]; C. Tsallis, Phys. Lett.
{\bf A206} 389 (1995). 

\bibitem{plastino} A.R. Plastino and A. Plastino, Phys. Lett. A {\bf 174}, 384
(1993); J.J. Aly, in "N-Body Problems and Gravitational Dynamics", Proc. of
the Meeting held at Aussois-France (21-25 March 1993), eds. F. Combes and
E. Athanassoula (Publications de l'Observatoire de Paris, 1993), p. 19; A.R.
Plastino and A. Plastino, Phys. Lett. A {\bf 193}, 251 (1994).

\bibitem{kaniadakis}G. Kaniadakis, A. Lavagno and P. Quarati, Phys. Lett.
B {\bf 369}, 308 (1996). 

\bibitem{lavagno}A. Lavagno, G. Kaniadakis, M. Rego-Monteiro, 
P. Quarati and C. Tsallis, {\it Non-extensive thermostatistical 
approach of the peculiar velocity function of galaxy clusters}, preprint 
(1996) [astro-phy 9607147].

\bibitem{hamity}V.H. Hamity and D.E. Barraco, Phys. Rev. Lett. {\bf 76},
4664 (1996). 

\bibitem{boghosian}B.M. Boghosian, Phys. Rev. E {\bf  53}, 4754 (1996); 
C. Anteneodo and C. Tsallis, {\it Two-dimensional turbulence in pure-electron 
plasma: A nonextensive thermostatistical description}, J. Mol. Liq. (1996), in press.

\bibitem{levy}P.A. Alemany and D.H. Zanette, Phys. Rev E {\bf 49}, 956
(1994); C. Tsallis, A.M.C. Souza and R. Maynard, in "L\'evy 
flights and related topics in Physics", eds. M.F. Shlesinger, G.M.
Zaslavsky and U. Frisch (Springer, Berlin, 1995), p. 269; D.H. Zanette
and P.A. Alemany, Phys. Rev. Lett. 75, 366 (1995); C. Tsallis, S.V.F. Levy,
A.M.C. Souza and R. Maynard, Phys. Rev. Lett. {\bf 75}, 3589 (1995) 
[Erratum: Phys. Rev. Lett. (1996), in press]; M.O. Caceres and C. E. 
Budde, Phys. Rev. Lett. {\bf 77}, 2589 (1996); D.H. Zanette and P.A. 
Alemany, Phys. Rev. Lett. {\bf 77}, 2590 (1996).

\bibitem{correlated}A.R. Plastino and A. Plastino, Physica A {\bf 222}, 347
(1995); C. Tsallis and D.J. Bukman, Phys. Rev. E {\bf 54}, R2197 (1996); 
A. Compte and D. Jou, J. Phys. A {\bf 29}, 4321 (1996).   

\bibitem{jund}P. Jund, S.G. Kim and C. Tsallis, Phys. Rev. B {\bf 52}, 
50 (1995); J.R. Grigera, Phys. Lett. A {\bf 217}, 47 (1996); S.A. Cannas and 
F.A. Tamarit, Phys. Rev. B {\bf 54}, R12661 (1996); L.C. Sampaio, M.P. 
de Albuquerque and F.S. de Menezes, {\it Nonextensivity and Tsallis 
statistics in magnetic systems}, Phys. Rev. B (1996), in press.

\bibitem{optimization}T.J.P. Penna, Phys. Rev. E {\bf 51}, R1 (1995) and
Computers in Physics {\bf 9}, 341 (1995); D.A. Stariolo and C. Tsallis,
Ann. Rev. Comp. Phys., vol. II, ed. D. Stauffer (World Scientific, Singapore,
1995), p. 343; K.C. Mundim and C. Tsallis, Int. J. Quantum Chem. {\bf 58},
373 (1996); J. Schulte, Phys. Rev. E {\bf 53}, 1348 (1996); I.
Andricioaei and  J. Straub, Phys. Rev. E {\bf 53}, R3055 (1996); C. Tsallis 
and D.A. Stariolo, 
Physica A {\bf 233}, 395 (1996); 
P. Serra, A.F. Stanton and S. Kais, {\it A new pivot method for global 
optimization}, Phys. Rev. E (1996), in press.

\bibitem{rajagopal}A.K. Rajagopal, Phys. Rev. Lett. {\bf 76}, 3469 (1996).

\bibitem{chaos} F.C. Moon, {\it Chaotic and Fractal Dynamics, An
Introduction for Applied Scientists and Engineers} (Wiley, New York, 1992);
E. A. Jackson, {\it Perspectives of Nonlinear Dynamics,
Vol. I and Vol. II} (Cambridge Univesrsity Press, New York, 1989, 1991).

\bibitem{definitions} J.D. Farmer, Z. Naturforsch {\bf37a}, 1304 (1982);
 G. Benetin, L.Galgani, and J.-M. Strelcyn, Phys. Rev. {\bf A14}, 2338
(1976); P. Grassberger and I. Procaccia, Phys. Rev. {\bf A28}, 2591
(1983); A.M.Ozorio de Almeida, {\it Hamiltonian Systems: Chaos and
Quantization}, Cambridge University Press (1988).

\bibitem{hilborn} R. C. Hilborn, {\it Chaos and Nonlinear Dynamics}
(Oxford University Press, New York, 1994), p. 390.

\bibitem{kolsinai} A.N. Kolmogorov, Dok. Acad. Nauk SSSR {\bf  119}, 861
(1958); Ya. G. Sinai, Dok. Acad. Nauk SSSR {\bf 124}, 768 (1959).  

\bibitem{tzp} C. Tsallis, A.R. Plastino, and W.-M. Zheng, {\it Power-law
Sensitivity to Initial Conditions - New Entropic Representations},
Chaos, Solitons and Fractals (1996), in press.

\bibitem{feigenbaum} M.J. Feigenbaum, J. Stat. Phys. {\bf 19}, 25 (1978).

\bibitem{hauser}P.R. Hauser, C. Tsallis and E.M.F. Curado, Phys. Rev. A 
{\bf 30}, 2074 (1984).

\bibitem{bailin} Hao Bai-lin, {\it Elementary Symbolic Dynamics} (World
Scientific, Singapore, 1989).

\bibitem{zanettekol}D.H. Zanette, Physica A {\bf 223}, 87 (1996).

\bibitem{beckscho} C. Beck and F. Schogl, {\it Thermodynamics of Chaotic
Systems: An Introduction} (Cambridge University Press, Cambridge, 1995).

\bibitem{grass} P. Grassberger, R. Badii and A. Politi, J. Stat. Phys. {\bf
51}, 135 (1988).

\bibitem{perbak}P. Bak, C. Tang and K. Wiesenfeld , Phys. Rev. A {\bf 38},
364 (1988); P. Bak and K. Chen,  Scientific American (January 1991), p.26. 

\end{references}
\end{document}